\documentclass[10pt,twocolumn,twoside]{IEEEtran}
\usepackage{fixltx2e}
\usepackage{inputenc}
\usepackage{amsmath}
\usepackage{amsfonts}
\usepackage{amssymb}
\usepackage{cite}
\usepackage{pict2e}
\usepackage{epsfig}
\usepackage{psfrag}
\usepackage{array}
\usepackage{verbatim}
\usepackage{threeparttable} 
\usepackage{algorithm}
\usepackage{algorithmic}
\usepackage{color}
\usepackage{subfigure}

\newtheorem{Theorem}{Theorem}

\newtheorem{remark}{Remark}
\newcommand{\rss}{\text{RSS}}

\newcommand{\bec}{}

\begin{document}
\IEEEoverridecommandlockouts
\title{Energy-Efficient Deterministic Adaptive Beamforming Algorithms for Distributed Sensor/Relay Networks$^\S$}
\author{\IEEEauthorblockN{Chun-Wei~Li$^*$, Kuo-Ming~Chen$^*$, Po-Chun~Fu, Wei-Ning~Chen, and Che Lin$^\dagger$}

\thanks{$^\S$
The work is supported by the National Science Council, R.O.C., under
grant number: MOST104-2221-E-007-041-MY2.}

\thanks{This work appeared in part in the Proceedings IEEE International Conference on Communications, Kyoto, Japan, June 5–9, 2011. }

\thanks{$^*$Equal contribution for both authors. 

$^\dag$Che Lin is the corresponding author.
Address: Institute of Communications Engineering, National Tsing Hua University, Hsinchu, Taiwan 30013, R.O.C.
E-mail: clin@ee.nthu.edu.tw.}

\thanks{
Chun-Wei Li, Kuo-Ming Chen, Po-Chun Fu and Wei-Ning Chen are with Institute of Communications Engineering, National Tsing Hua University, Hsinchu, Taiwan 30013, R.O.C. 
E-mail: 
\{re21962000, wxes9250605, nctuce.fu, iamlingo080990 \}@gmail.com
}

}
\maketitle
\noindent 

\begin{abstract}
In this study,  energy-efficient deterministic adaptive beamforming algorithms are proposed for distributed sensor/relay networks. Specifically, DBSA, D-QESA, D-QESA-E, and a hybrid algorithm, hybrid-QESA, that combines the benefits of both deterministic and random adaptive beamforming algorithms, are proposed. 
Rigorous convergence analyses are provided for all our proposed algorithms and convergence to the global optimal solution is shown for all our proposed algorithms. Through extensive numerical simulations, we demonstrate that superior performance is achieved by our proposed DBSA and D-QESA over random adaptive beamforming algorithms for static channels. Surprisingly, D-QESA is also more robust against random node removal than random adaptive beamforming algorithms. For time-varying channels, hybrid-QESA indeed achieves the best performance since it combines the benefits of both types of adaptive beamforming algorithms. In summary, our proposed deterministic algorithms demonstrate superior performance both in terms of convergence time and robustness against channel and network uncertainties. 
\end{abstract}
\textbf{Index Terms - Beamforming, convergence analysis, distributed algorithms, time-varying channels}


\section{Introduction}\label{sec:intro}
There has been a demand for improving the reliability and energy efficiency of wireless communication systems pertaining to distributed sensor/relay networks. A lack of constant power supply in a distributed wireless system, such as a wireless sensor or a relay network, and the possibility of hardware failures have driven the emergence of energy-efficient algorithms for such applications. In particular, distributed beamforming has been proposed as a viable solution where all distributed transmitters seek to align in phase at the receiver end. However, it is difficult to implement such transmit beamforming scheme in a distributed fashion in practice because perfect channel state information (CSI) needs to be made available at all distributed transmitters. Tremendous overhead is required to feed back CSI from the receiver to all distributed transmitters. 

In literature, the efforts of designing efficient distributed adaptive beamforming algorithms can be categorized into two classes: random adaptive beamforming algorithms or deterministic adaptive beamforming algorithms. 
The latter has been greatly investigated in the literature over the last decade. In \cite{1}, a randomized approach was proposed in which the receiver uses a feedback
link that allows each transmitter to make proper phase adjustment. In \cite{2}, Mudumbai et al investigated an adaptive distributed beamforming paradigm that requires only a single bit feedback. The scalability and the rate of convergence analysis of that scheme were given in \cite{3}. By considering this algorithm as a local random search algorithm, \cite{4} provided a comprehensive analysis of the fast convergence of the algorithm. An extensive
convergence analysis of these scheme was presented in \cite{5}.
A new scheme inspired from a completely different field
was proposed in \cite{6}, where a bio-inspired robust adaptive
random search algorithm was presented, and has been proven
to converge in probability. Other examples of feedback-based synchronization procedures include \cite{7, 8, 9, 10}.

Relative few efforts have been spent on the studies of deterministic adaptive beamforming algorithms. Several deterministic algorithms for distributed beamforming have  been proposed in the literature. For example, Thibault et al introduced a deterministic algorithm with individual power constraint \cite{12}. This algorithm was extended to the case of time-varying channels in \cite{13}. Simulation results show that the deterministic approaches outperform the random methods in the case of static and time-varying channels. For amplify-and-forward wireless relay networks, an algorithm using additive deterministic perturbations was presented in \cite{14}. Fertl et al further investigated a multiplicative deterministic perturbations for distributed beamforming under a total power constraint \cite{15}. Although faster convergence is often observed for deterministic adaptive beamforming algorithms, they are more sensitive to channel uncertainties in general.

In this study, we investigate the design and analysis of deterministic adaptive beamforming algorithms. 
We propose a Deterministic Bisection Search Algorithm (DBSA) that is inspired by the idea of bisection search. 
Furthermore, we propose a Deterministic Quadratic Equation Search Algorithm (D-QESA), inspired by a shift of perspective that views each transmission as a RSS function evaluation and views the problem of adaptive beamforming for each distributed transmitter as solving quadratic equations with independent variables. 
For the case of equal channel gains, we propose a modified version of D-QESA, termed D-QESA-E, to speed up convergence for this setting.
Note that deterministic algorithms usually have worse performance for time-varying channels. 
Therefore, we further propose a hybrid algorithm, hybrid-QESA, that combines the benefits of both deterministic and random adaptive beamforming algorithms.

The remaining part of this paper is organized as follows. In Section II, we presents a system model for the adaptive distributed beamforming problem and formulate the phase alignment problem. The details of DBSA is described in Section III. 
In Section IV, we propose D-QESA, D-QESA-E, and hybrid-QESA. 
Convergence analysis is conducted for all our proposed algorithms in Section V. 
Extensive simulation results are presented in Section VI to demonstrate the performance advantages of our proposed algorithms. We conclude the paper in Section VII.

\section{System Model and Problem Formulation}
In this study, we consider a wireless sensor/relay network consisting of distributed transmitters where  a common message $ s \in C$ is to be conveyed to the receiver end. Let $N_s$ be the total number of distributed transmitters, each with an average power constraint of $E[|s|^2]$. Each transmitter and the receiver is assumed to be equipped with single antenna. We assume that the channel between them is frequency flat and slow fading. In addition, there exists an error-free feedback link from the receiver to all distributed transmitters.

According to these system specifications, the discrete-time, complex baseband model is given by 
\begin{equation}\label{eq:model}
\begin{aligned}
y[n] &= \sum_{i=1}^{N_s} h_i[n] g_i[n] s + w[n] \\&= \sum_{i=1}^{N_s} a_i[n] b_i[n] e^{j(\phi_i[n] + \psi_i[n])} s + w[n]
\end{aligned}
\end{equation}
where $y[n]$ denotes the received signal, $h_{i}[n] = a_i[n] e^{j \phi_i[n]}$ corresponds to the channel coefficient, $g_{i}[n] = b_{i}[n]e^{j \psi_{i}[n]} $ corresponds to the beamforming coefficient, and $w[n]$ is the additive white Gaussian noise for the $i$-th distributed transmitter.

To simplify notation, let $\theta_i[n]=\phi_i[n]+\psi_i[n]$ be the total phase of received signal from $i$-th transmitter during the $n$-th transmission. Also, we assume that $s=\sqrt{P}$ and impose a fixed power constraint $b_i=1$ among transmitters. Therefore, the received signal can be expressed as $\sqrt{P}\sum_{i=1}^{N_s}a_i[n]e^{j\theta _i [n]}$. Finally, we assume that the strength of the composite signal from distributed transmitters can be perfectly estimated at the destination and the {received signal strength (RSS)} function can be described as 
\begin{equation}\label{eq:mag}
\rss(\theta_1[n],\cdots,\theta_{N_s}[n]) = \sqrt{P} \left | \sum_{i=1}^{N_s} a_i[n] e^{j\theta_i[n]} \right |
\end{equation}

Note that the primary goal here is to maximize the RSS function given in (\ref{eq:mag}) for distributed beamforming schemes to exploit the potential power gain efficiently. Furthermore, it is clear that to reach the global maximum, all phases need to be completely aligned, i.e, $\theta_1[n]=\theta_2[n]=\cdots =\theta_{N_s}[n]$. Here,  we denote $\rss_{max}[n]$ as the global maximum value of the RSS function, i.e.,
\begin{equation}
\rss_{max}[n] = \sqrt{P} \left | \sum_{i=1}^{N_s} a_i[n] \right |
\end{equation}

We then denote the beamforming gain ratio $\rho_{N_{s}}[n]$  as

\begin{equation}
\rho_{N_{s}}[n]=\dfrac{\rss(\theta_1[n],\cdots,\theta_{N_s}[n])}{\rss_{max}[n] }
\end{equation}

The beamforming gain ratio represents how close to the global maximum the considered algorithm reaches, and attains its maximum value of $1$ when perfect phase alignment is achieved.
Note that $\rss_{max}[n]$ and $\rho_{N_{s}}[n]$ can vary with time if we consider uncertainties in the wireless network topology or time-varying channel coefficients.

\section{Deterministic Biscetion Search Algorithm}
In this section, we introduce in details the proposed DBSA algorithm, which is inspired by the idea of bisection search. To simplify notation, let us assume that the phase of each distributed transmitter remains confined within an angle varying betweens $[0,2\pi]$. By viewing the phase alignment problem as a maximization problem within an $N_s$ dimensional hypercube with an edge length of $2\pi$, it is then suggested that the searching process for the global maximum can be done in a bisection fashion.

The phase alignment procedure is done in a greedy fashion within the proposed DBSA. In fact, only one distributed transmitter is permitted to adjust its phase at any time while all other transmitters keep their phase unchanged. With the pseudocode given in Algorithm 1, we elaborate the algorithm by providing a detailed step-by-step description as follows:

\textbf{Step 1- Initialization:}
The system is initialized by randomly generating the initial phase $\theta [0]=\{\theta_{1}[0],\cdots,\theta_{N_s}[0]\}\in \mathbb{R}^{N_s}$. The phase adjustment parameter $\alpha$, known to all transmitters, should be configured to a number $\alpha_{init}$ that divides $2\pi$. Distributed transmitters jointly beamform a fixed symbol $s=\sqrt{P}$ using the generated phase configuration. The receiver then records the value of the RSS function given by (\ref{eq:mag}) corresponding to the initial phase $\theta[0]$.

\textbf{Step 2- Initial Phase Rotation:}
The algorithm undergoes $N_{s}$ rounds of phase evaluations corresponding to the $N_{s}$ distributed transmitters.

During the $i$-th round, only the $i$-th transmitter transmits for $K-1$ iterations. The $i$-th transmitter alters its phase by iterating through the elements in the set $\Theta_{i}$ given by:
\begin{equation}
\Theta_{i}=\{\theta_{i}[0]+\alpha, \cdots, \theta_{i}[0]+(K-1)\alpha\}\ \forall i\leq N_{s}
\end{equation}
where $K$ is an integer that equals to $2\pi/\alpha$. At each iteration, the receiver compares the RSS function with the largest RSS function recorded so far, and feedback a single bit that indicates whether the current selected phase configuration by the $i$-th transmitter exceeds previous recorded RSS function. Should the RSS function exceed the largest recorded value, the feedback bit is set to 1, and to 0 otherwise.

By the time all phase configurations in $\Theta_{i}$. are tested, the$i$-th transmitter should be able to distinguish the phase configuration that correspond to the best RSS performance, and this particular phase configuration should replace its initial phase configuration generated in \textbf{Step 1}, i.e., the $i$-th element of $\theta [0]$. Once all transmitters are done with the evaluation procedures, the parameter $\alpha$ is halved, and the algorithm enters \textbf{Step 3}.

\textbf{Step 3- Forward Adjustment:}
Before proceeding to \textbf{Step 5}, the algorithm undergoes $N_{s}$ iterations where each iteration involves procedures \textbf{Step 3} and \textbf{Step 4}. At the $n$-th iteration, the $n$-th distributed transmitter adjusts its phase by adding up the parameter $\alpha$ to its phase, while all other distributed transmitters keep their phases unchanged. From the system point of view, the adjusted phase vector can be expressed as:

\begin{equation}
\begin{aligned}
{\boldsymbol\theta}^{\prime}[n]&=[\theta_{1}[n-1], \cdots, \theta_{n-1}[n-1], \theta_{n}[n-1]+\alpha,  \\
&~~~\theta_{n+1}[n-1], \cdots, \theta_{N_{s}}[n-1]]^{T} 
\end{aligned}
\end{equation}

The receiver compares the RSS function corresponding to the adjusted phase vector $\theta^{'}[n]$ with the highest recorded RSS function up to the current iteration, and broadcasts a one-bit feedback signal ${\cal B}_{F}$ back to all transmitters. If improvement on the RSS function is observed, the adjusted phase is kept by the $n$-th transmitter and the algorithm proceeds to the next iteration that adjusts the $(n+1)$-th transmitter provided that $n< N_s$. Otherwise, if no RSS improvement is observed, the algorithm enters \textbf{Step 4}, or if $n=N_s$, the algorithm enters \textbf{Step 5}.\\

More precisely, the feedback signal ${\cal B}_{F}$ is generated as
\begin{equation}
\begin{aligned}\label{eq:feedback signal}
{\cal B}_{F}=1\{{\rm \rss}(\theta^{\prime}[n])>\max_{i=1, \cdots, n-1}{\rm \rss}(\theta[i])\}  
\end{aligned}
\end{equation}

\textbf{Step 4- Reverse Adjustment}
At this stage of the algorithm, the $n$-th distributed transmitter adjusts its phase by adding the inverse of the adjustment parameter to its phase, where the inversely adjusted phase vector $\theta^{\prime \prime}[n]$ can be expressed as
\begin{equation}
\begin{aligned}
\mathbf{\boldsymbol{\theta}^{\prime \prime}}[n]&=[\theta_{1}[n-1], \cdots, \theta_{n-1}[n-1], \theta_{n}[n-1]-\alpha,  \\
&~~~\theta_{n+1}[n-1], \cdots, \theta_{N_{s}}[n-1]]^{T} 
\end{aligned}
\end{equation}

As in the previous step, the receiver compares the RSS function corresponding to $\theta^{\prime \prime}[n]$ with the largest recorded RSS function and broadcasts a one-bit feedback signal ${\cal B}_{R}$ back to transmitters. Upon receiving "$1$" from the receiver, the $n$-th transmitter updates its phase with the inversely adjusted value, while no action is taken for the cases of feedback being "$0$". The feedback signal ${\cal B}_{R}$ is generated as in (\ref{eq:feedback signal}) with $\theta^{\prime}[n]$ replaced with $\theta^{\prime \prime}[n]$, and the procedures throughout \textbf{Step 3} and \textbf{Step 4} can be characterized by the following expression:

\begin{equation}
\begin{aligned}
\theta[n]=\left\{\begin{array}{c}
\theta^{\prime}[n], ~~~{\cal B}_{F}=1~~~~~~~~~~~\\ 
\theta^{\prime\prime}[n], ~~~{\cal B}_{F}=0, {\cal B}_{R}=1 \\ \theta[n-1],  ~~~ o.w.~~~~~~~~~~~
\end{array}\right..
\end{aligned}
\end{equation}

Once completed with the updating process, the algorithm returns to \textbf{Step 3} and proceed to the $(n+1)$-th iteration where the phase of the $(n+1)$-th transmitter can be adjusted, or, if $n=N$, the algorithm enters \textbf{Step 5}.

\begin{algorithm}[h] \label{alg: DBSA for static channels}
\caption{Deterministic Biscetion Search Algorithm (DBSA)} 
Initialize parameters

$\textbf{repeat}$

~~~~$k=0$

~~~~$\textbf{for}$ $i\leftarrow 1$ to $N_s$ $\textbf{do}$

~~~~~~~~$\textbf{for}$ $j\leftarrow 1$ to $\lfloor (360/\alpha)-1\rfloor$ $\textbf{do}$

~~~~~~~~~~~~$\theta_i^{\prime}=\theta_i[0]+\alpha \cdot j$

~~~~~~~~~~~~$\textbf{if}$ $\text{\rss}(\theta^{\prime}[0])>\text{\rss}(\theta[0])$ $\textbf{then}$ 

~~~~~~~~~~~~~~~~$\theta[0]\leftarrow \theta^{\prime}[0]$

~~~~~~~~~~~~$\textbf{end if}$

~~~~~~~~$\textbf{end for}$

~~~~$\textbf{end for}$

~~~~$\textbf{for}$ $i\leftarrow 1$ to $N_s$ $\textbf{do}$

~~~~~~~~$k=k+1$

~~~~~~~~$\theta_i[k]\leftarrow \theta_i[k-1]+\alpha$

~~~~~~~~$\textbf{if}$ $\text{\rss}(\theta[k])<\text{\rss}(\theta[k-1])$ $\textbf{then}$ 

~~~~~~~~~~~~$\theta_i[k]\leftarrow \theta_i[k-1]-\alpha$

~~~~~~~~~~~~$\textbf{if}$ $\text{\rss}(\theta[k])>\text{\rss}(\theta[k-1])$ $\textbf{then}$ 

~~~~~~~~~~~~~~~~$\theta[k]\leftarrow \theta[k-1]$

~~~~~~~~~~~~$\textbf{end if}$

~~~~~~~~$\textbf{end if}$

~~~~$\textbf{end for}$

$\alpha\leftarrow\dfrac{\alpha}{2}$

$\textbf{until}$ stopping criteria reached

\end{algorithm}

\textbf{Step 5- Parameter Adjustment:}
Upon entering \textbf{Step 5}, all transmitters scale down the phase adjustment parameter $\alpha$ to half of its current value, and return to \textbf{Step 3}. For notational simplicity, the iteration index is reset to $1$ and $\theta [0]$ is replaced with $\theta [N_s]$.

\textbf{Stopping criterion:}
Given a threshold value for the RSS function, if the RSS function is greater than or equal to this target value at any step of the algorithm, the phase alignment process is said to be completed. The threshold value may be based on any statistical information and has to be available and known to the receiver.

\section{Deterministic Quadratic Equation Search Algorithm}

Our proposed D-QESA algorithm undergoes $N_s$ rounds of RSS function evaluations corresponding to the $N_s$ distributed transmitters.
During the $i$-th round, only the $i$-th transmitter transmits for $2$ additional iterations. We will describe how these $2$ additional RSS function evaluations can be used to improve phase alignment for different wireless environment settings in the following subsections. To begin with, note that the RSS function in (\ref{eq:mag}) can be rewritten as 

\begin{eqnarray}
&\rss(\theta_1[n],\cdots,\theta_{N_s}[n]) = \sqrt{P} \left | \sum_{i=1}^{N_s} a_i[n] e^{j\theta_i[n]} \right | \\
& = \sqrt{P} \left | \sum_{\substack{k=1 \\ k\neq i}}^{N_s} a_k[n] e^{j\theta_k[n]} + a_i[n] e^{j\theta_i[n]} \right | \\
& = \sqrt{P} \left | {r}_i + {t}_i  \right |\label{eq:Substitution} \\
& = \sqrt{P} \sqrt{\left(\left| r_i \right| + \left| t_i\right|\cos\beta_i\right)^2+\left( \left| t_i\right|\sin\beta_i \right)^2}\label{eq:Substitution 1}
\end{eqnarray}
where ${r}_i=\sum_{ k\neq i} a_k[n] e^{j\theta_k[n]}$, ${t}_i=a_i[n] e^{j\theta_i[n]}$ and $\beta_i$ is the phase angle between complex numbers $r_i$ and $t_i$. Note that the time indices of $\beta_i$, $r_i$, and $t_i$ are omitted for notation simplicity.
Furthermore, if one can only adjust the beamforming phase $\theta_i[n]$ for each iteration, the optimal strategy would be aligning the direction of $r_i$ and $t_i$, i.e., rotating $\theta_i[n]$ to cancel out the phase angle $\beta_i$. This is the main idea behind D-QESA.

\subsection{D-QESA for static channels}

The pseudocode of the proposed algorithm can be found in Algorithm 2. For static channels, the channel coefficients do not change with time, i.e., $h_{i} = a_i e^{j \phi_i}$. Under this setting, we elaborate the proposed D-QESA in a step-by-step fashion as follows.

\textbf{Step 1- Initialization:}
The system is initialized by randomly generating the initial phase $\theta [0]=\{\theta_{1}[0],\cdots,\theta_{N_s}[0]\}\in \mathbb{R}^{N_s}$. In the beginning, we set $\theta_{cur,i} =\theta [0]$ where $\theta_{cur,i}$ represents the current phase of the $i$-th distributed transmitter. $\theta_{temp}$ represents the temporary value of phase. The parameter $n$ is the time index with an initial value of 0. The phase adjustment parameters $\alpha$ and  $\eta$ are known to all transmitters. The angles $\alpha$ and $\eta$ are  initialized as $\alpha=\pi$ and $\eta=\pi /2$. 

As mentioned earlier, the aim of D-QESA is to compute the value of  $\beta_i$ based on $2$ additional RSS function evaluations and feed it back to the $i$-th transmitters to adjust its phase.
In the following steps, we elaborate on how this is done based on only $2$ additional RSS function evaluations in each round.

\textbf{Step 2- The phase rotation of $\alpha$ and $\eta$:}
Denote the initial RSS function by $M_1$. We first rotate the phase of the $i$-th transmitter by $\alpha$ to obtain an updated RSS function value of $M_2$. If $M_2>M_1$, ${\cal B}_{F}$ is set to $1$ and is fed back to distributed transmitters. The $i$-th transmitter then keeps this adjusted phase since the RSS function is improved. Otherwise,  ${\cal B}_{F}=0$ is fed back and the $i$-th transmitter reverts back to its original phase. Note that this step not only improve the RSS function but also make  $\beta_i$, to be solved later, falls in $[-\pi/2, \pi/2]$, the benefit of which will become clear later.


Next, we obtain the second RSS function evaluations by
rotating the phase of the $i$-th transmitter by $\eta$. Through this step, we obtain an updated RSS function value of $M_3$ at the receiver. 

\begin{algorithm}[h] \label{alg: DDSA for static channels}
\caption{Deterministic Quadratic Equation Search Algorithm (D-QESA)} 
Initialize parameters

$\textbf{repeat}$

~~~~$\textbf{for}$ $i\leftarrow 1$ to $N_s$ $\textbf{do}$

~~~~~~~~$\text{test}\leftarrow 0$

~~~~~~~~$n\leftarrow n+1$

~~~~~~~~$\theta_{temp}\leftarrow \theta_{cur,i}$

~~~~~~~~$M_{1}\leftarrow \rss_{temp}(\theta_{temp})$

~~~~~~~~$\rss[n]\leftarrow M_1$

~~~~~~~~$n\leftarrow n+1$

~~~~~~~~$\theta_{temp}\leftarrow \theta_{cur,i}+\alpha$

~~~~~~~~$M_{2}\leftarrow \rss_{temp}(\theta_{temp})$

~~~~~~~~$\textbf{if}$ $M_{2}>\rss[n-1]$ $\textbf{then}$ 

~~~~~~~~~~~~$\rss[n]\leftarrow M_{2}$

~~~~~~~~~~~~$\theta_{cur,i}\leftarrow \theta_{cur,i}+\alpha$

~~~~~~~~~~~~$\text{test}\leftarrow 1$

~~~~~~~~~~~~receiver feeds back a 1 bit to transmitter

~~~~~~~~$\textbf{else}$

~~~~~~~~~~~~$\rss[n]\leftarrow \rss[n-1]$

~~~~~~~~~~~~receiver feeds back a 0 bit to transmitter

~~~~~~~~$\textbf{end if}$

~~~~~~~~$\textbf{if}$ $n\leq 3N$ $\textbf{then}$ 

~~~~~~~~~~~~$n\leftarrow n+1$

~~~~~~~~~~~~$\textbf{if}$ $\text{test}=1$ $\textbf{then}$ 

~~~~~~~~~~~~$\theta_{temp}\leftarrow \theta_{cur,i}+\eta-\alpha$

~~~~~~~~~~~~$\textbf{else}$

~~~~~~~~~~~~~~~~$\theta_{temp}\leftarrow \theta_{cur,i}+\eta$

~~~~~~~~~~~~$\textbf{end if}$

~~~~~~~~~~~~$M_3 \leftarrow \rss_{temp}(\theta_{temp})$\\

~~~~~~~~~~~~$x\leftarrow(M_{1}^2+M_{2}^2)/2$\\

~~~~~~~~~~~~$\beta_i\leftarrow\arctan\left[ \dfrac{(M_3^2-x)}{(M_2^2-x)}\right]$\\

~~~~~~~~~~~~$|t_i|\leftarrow \sqrt{\dfrac{x-\sqrt{x^2-\dfrac{(M_2^2-x)^2}{(\cos\beta_i)^2}}}{2P}}$

~~~~~~~~~~~~$\theta_{cur,i}\leftarrow \theta_{cur,i}+\beta_i$

~~~~~~~~$\textbf{else}$

~~~~~~~~~~~~$\beta_i\leftarrow\arccos\left( \dfrac{M_1^2-M_2^2}{2|t_i|\sqrt{2(M_1^2+M_2^2)P-4|t_i|^2P^2}}\right)$\\

~~~~~~~~~~~~$\theta_{cur,i}\leftarrow \theta_{cur,i}+\beta_i$

~~~~~~~~$\textbf{end if}$

~~~~$\textbf{end for}$

$\textbf{until}$ stopping criteria reached

\end{algorithm}

\textbf{Step 3- Calculate the value of $\beta_i$ and $|t_i|$:}
Note that $M_1$ represents the value of the initial RSS function value. $M_2$ and $M_3$ were also obtained through $2$ additional RSS function evaluations. Based on (\ref{eq:Substitution}) , we can obtain
\begin{equation}\label{eq:M1,M2,M3}
\begin{aligned}
M_1^2&=P\left(   |r_i |^2+ |t_i |^2+2 |r_i | |t_i |\cos\left(\beta_i\right)\right) \\
M_2^2&=P\left(   |r_i |^2+ |t_i |^2+2 |r_i | |t_i |\cos\left(\beta_i+\alpha\right)\right) \\
M_3^2&=P\left(   |r_i |^2+ |t_i |^2+2 |r_i | |t_i |\cos\left(\beta_i+\eta\right)\right) 
\end{aligned}
\end{equation}

Note that we describes $3$  quadratic equations with $3$ independent variable $|r_i|$, $|t_i|$ and $\beta_i$. Therefore, we can obtain $|r_i|$, $|t_i|$ and $\beta_i$ by solving these equations. 
Specifically, let $x=(M_{1}^2+M_{2}^2)/2$, we can obtain 
\begin{eqnarray}
x=P\left( |r_i |^2+ |t_i |^2\right) 
\end{eqnarray}

Then,
\begin{equation}\label{calculate}
\begin{aligned}
M_{2}^2-x&=2P|r_i | |t_i |\cos\left(\beta_i +\pi \right) \\
M_{3}^2-x&=2P|r_i | |t_i |\cos\left(\beta_i +\dfrac{\pi}{2}\right)\\
\Rightarrow \dfrac{M_{3}^2-x}{M_{2}^2-x}&=\dfrac{\cos(\beta_i +\dfrac{\pi}{2})}{\cos\left(\beta_i +\pi\right)}=\dfrac{-\sin(\beta_i)}{-\cos(\beta_i)}=\tan(\beta_i)
\end{aligned}
\end{equation}

Now, we can obtain $\beta_i$ by 
\begin{equation}
\begin{aligned}
\beta_i= \arctan\left( \dfrac{M_{3}^2-x}{M_{2}^2-x}\right) 
\end{aligned}
\end{equation}

It is clear that $|t_i|=a_i$ can be obtained by plugging (\ref{calculate}) into (\ref{eq:Substitution 1}). Specifically, we obtain

\begin{equation}\label{eq:ti}
\begin{aligned}
|t_i|=\sqrt{\dfrac{x-\sqrt{x^2-\dfrac{(M_2^2-x)^2}{(\cos\beta_i)^2}}}{2P}}
\end{aligned}
\end{equation}

\textbf{Step 4- With the information of $|t_i|$, adjust the phase in the second round-robin and after:}
During the first round-robin, we calculate the values of $\beta_1$ and $|t_i|$ for all $i$. 
Note that one can simply calculate the value of $\beta_i$ and feed it back to adjust the phase of the $i$-th transmitter. Obtaining the value of $|t_i|$ seems irrelevant for the current round. However, it is important to note that $|t_i|$ equals the magnitude of the channel gain $a_i$ from the $i$-th distributed transmitter to the receiver and does not vary with time for a static channel. Furthermore, we run D-QESA in a round-robin fashion until the algorithm converges. 
By calculating and keeping the value of $|t_i|$, we can reduce the $2$ additional RSS function evaluations into only $1$ since now we only have two unknown independent variables and only $M_1$ and $M_2$ are needed when we reaches the second round-robin and after. Specifically, from (\ref{eq:M1,M2,M3}), we have 

\begin{equation}\label{eq:M1 M2}
\begin{aligned}
M_1^2-M_2^2=4P|r_i||t_i|\cos(\beta_i)\\
M_1^2+M_2^2=2P|r_i|^2+2P|t_i|^2\\
\end{aligned}
\end{equation}

Then, we can derive

\begin{equation}
\begin{aligned}
|r_i|=\sqrt{\dfrac{M_1^2+M_2^2}{2P}-|t_i|^2}
\end{aligned}
\end{equation}

Plugging (\ref{eq:M1 M2}) into (\ref{eq:ti}), we obtain
\begin{equation}
\begin{aligned}
M_1^2-M_2^2=4P\left(\sqrt{\dfrac{M_1^2+M_2^2}{2P}-|t_i|^2}\right)|t_i|\cos(\beta_i)
\end{aligned}
\end{equation}

Then, we can solve for $\beta_i$ to obtain
\begin{equation}
\begin{aligned}
\beta_i =\arccos\left( \dfrac{M_1^2-M_2^2}{2|t_i|\sqrt{2(M_1^2+M_2^2)P-4|t_i|^2P^2}}\right) 
\end{aligned}
\end{equation}

This extra information of $|t_i|$ allows us to cut down the convergence time by about $1/3$ after the second round-robin, which is a significant improvement.

Once $\beta_i$ is obtained, we can predict and adjust the phase at the $i$-th distributed transmitter accordingly. Ideally, one can feed back $\beta_i$ directly to the $i$-th distributed transmitter to achieve perfect phase alignment from the viewpoint of the $i$-th transmitter. However, if there is only limited bandwidth for the reverse feedback link ${\cal B}_{F}$, proper quantization is necessary. For example, if $2$ bits are available for the reverse feedback link, $\beta_i$ can be quantized as $\{ -3\pi/8, -\pi/8, \pi/8, 3\pi/8 \}$. With this feedback information, the $i$-th transmitter can decide whether to subtract either one of $\{ -3\pi/8, -\pi/8, \pi/8, 3\pi/8 \}$ to achieve a higher value of the RSS function. Note that if $\beta_i$ is 0 or close to 0, no information is fed back to distributed transmitters and the phase of the $i$-th transmitter is left unchanged.

\textbf{Stopping criterion:}
Given a threshold value for the RSS function, if the achieved value of the RSS function is greater than or equal to this target value at any step of the algorithm, the phase alignment process is said to be completed. The threshold value may be set based on the statistical information of channels.

\subsection{D-QESA-E for equal channel gains}

For the case where we have equal gains for all channels, i.e., $a_{i}=a,~\forall i$, further modifications can be made to  speed up our proposed D-QESA. We term the modified algorithm as Deterministic Quadratic Equation Search Algorithm - Equal channel gains (D-QESA-E). 
The pseudocode of the proposed algorithm can be found in Algorithm 3. Details of the algorithm is elaborated in the following steps.


\textbf{Step 1- Initialization:}
The system is initialized by randomly generating the initial phase $\theta [0]=\{\theta_{1}[0],\cdots,\theta_{N_s}[0]\}\in \mathbb{R}^{N_s}$. In the beginning, we set $\theta_{cur,i} =\theta [0]$ where $\theta_{cur,i}$ represent the current phase of the $i$-th transmitter. Again, the parameters $\alpha$ and $\eta$ are  initialized to $\alpha=\pi$ and $\eta=\pi /2$. As in D-QESA, $\beta_i$ is the value that we seek to calculate and feed back to the $i$-th transmitters for phase adjustment. The value of the RSS function for this initial phase is again denoted by $M_1$.

\textbf{Step 2- The first phase rotation of $\alpha$ and $\eta$:}
Here, the procedure is same as that in Algorithm 2. First, we rotate the phase by $\alpha$ and use a one-bit feedback to determine whether the phase rotation is beneficial or not. If so, the updated phase is kept.  Otherwise, the original phase is used. An updated RSS function value is obtained and denoted as $M_2$. Then, we rotate the phase by $\eta$ to obtain another RSS function evaluation $M_3$. Note that this step is only done once. After the first round, only one additional RSS function evaluation is necessary for reasons that will become clear later.

\textbf{Step 3- Calculate the value of $\beta_i$ and $|t_i|$:}
The calculation of $\beta_i$ and $\left|t_i \right|$ is almost the same as before. There is, however, an important difference. For the case of equal channel gains, $a_i=a,~ \forall i$. This implies that $|t_i|=|a_i e^{j\theta_i[n]}|=a$. After the first round of phase adjustment, we obtain $|t_1|=a$ and therefore, only two independent variables $\beta_i$ and $|r_i|$ remain for the ensuing rounds. This important observation suggests that only $M_1$ and $M_2$ are needed for the calculation of $\beta_i$. That is, 

\begin{equation}
\begin{aligned}
\beta_i=\arccos\left( \dfrac{M_1^2-M_2^2}{2|t_i|\sqrt{2(M_1^2+M_2^2)P-4|t_i|^2P^2}}\right) 
\end{aligned}
\end{equation}

This modification allows us to cut down the convergence time of our proposed algorithm by almost $1/3$. This is a significant improvement for energy-efficient algorithms in distributed wireless sensor/relay networks.  

\textbf{Stopping criterion:}
Given a threshold value for the RSS function, if the value of the RSS function is greater than or equal to this target value at any step of the algorithm, the phase alignment process is said to be completed. 

\begin{algorithm}[h] \label{alg: DDSA for static channels for same a}
\caption{Deterministic Quadratic Equation Search Algorithm - Equal channel gains (D-QESA-E)} 
Initialize parameters

$\textbf{repeat}$

~~~~$\textbf{for}$ $i\leftarrow 1$ to $N_s$ $\textbf{do}$

~~~~~~~~$\text{test}\leftarrow 0$

~~~~~~~~$n\leftarrow n+1$

~~~~~~~~$\theta_{temp}\leftarrow \theta_{cur,i}$

~~~~~~~~$M_{1}\leftarrow \rss_{temp}(\theta_{temp})$

~~~~~~~~$\rss[n]\leftarrow M_1$

~~~~~~~~$n\leftarrow n+1$

~~~~~~~~$\theta_{temp}\leftarrow \theta_{cur,i}+\alpha$

~~~~~~~~$M_{2}\leftarrow \rss_{temp}(\theta_{temp})$

~~~~~~~~$\textbf{if}$ $M_{2}>\rss[n-1]$ $\textbf{then}$ 

~~~~~~~~~~~~$\rss[n]\leftarrow M_{2}$

~~~~~~~~~~~~$\theta_{cur,i}\leftarrow \theta_{cur,i}+\alpha$

~~~~~~~~~~~~$\text{test}\leftarrow 1$

~~~~~~~~~~~~receiver feeds back a 1 bit to transmitter

~~~~~~~~$\textbf{else}$

~~~~~~~~~~~~$\rss[n]\leftarrow \rss[n-1]$

~~~~~~~~~~~~receiver feeds back a 0 bit to transmitter

~~~~~~~~$\textbf{end if}$

~~~~~~~~$\textbf{if}$ $n= 2$ $\textbf{then}$ 

~~~~~~~~~~~~$n\leftarrow n+1$

~~~~~~~~~~~~$\textbf{if}$ $\text{test}= 1$ $\textbf{then}$ 

~~~~~~~~~~~~~~~~$\theta_{temp}\leftarrow \theta_{cur,i}+\eta-\alpha$

~~~~~~~~~~~~$\textbf{else}$

~~~~~~~~~~~~~~~~$\theta_{temp}\leftarrow \theta_{cur,i}+\eta$

~~~~~~~~~~~~$\textbf{end if}$

~~~~~~~~~~~~$M_3 \leftarrow \rss_{temp}(\theta_{temp})$

~~~~~~~~~~~~$x\leftarrow(M_{1}^2+M_{2}^2)/2$

~~~~~~~~~~~~$\beta_i\leftarrow\arctan\left[ \dfrac{(M_3^2-x)}{(M_2^2-x)}\right]$

~~~~~~~~~~~~$|t_i|\leftarrow \sqrt{\dfrac{x-\sqrt{x^2-\dfrac{(M_2^2-x)^2}{(\cos\beta_i)^2}}}{2P}}$

~~~~~~~~~~~~$\theta_{cur,i}\leftarrow \theta_{cur,i}+\beta_i$

~~~~~~~~$\textbf{else}$

~~~~~~~~~~~~$\beta_i\leftarrow\arccos\left( \dfrac{M_1^2-M_2^2}{2|t_i|\sqrt{2(M_1^2+M_2^2)P-4|t_i|^2P^2}}\right)$

~~~~~~~~~~~~$\theta_{cur,i}\leftarrow \theta_{cur,i}+\beta_i$

~~~~~~~~$\textbf{end if}$

~~~~$\textbf{end for}$

$\textbf{until}$ stopping criteria reached

\end{algorithm}

\subsection{Hybrid-QESA for time-varying channels }
In this subsection, we examine the RSS function of D-QESA under time-varying environments. Here, we assume that the time-varying channel phase is a one-step Markov process, i.e., 
\begin{equation}\label{time-varying}
\phi_i [n]=\phi_i [n-1]+\xi_i [n]
\end{equation}
where the sequence of $\xi_i[n]$ for $i=1, 2, \ldots, N_s$ consists of i.i.d. Gaussian noise. Namely, $\xi_i[n] \sim N(0,\sigma_{\xi}^2)$. 

For time-varying channels, deterministic adaptive beamforming algorithms are worse than their random counterpart in general. Therefore, a hybrid algorithm is proposed that combines the advantages of both kinds of adaptive beamforming algorithms. We term the proposed hybrid algorithm as hybrid-QESA.
In hybrid-QESA, we initialize the algorithm by running D-QESA for an entire round-robin, i.e., each distributed transmitter updates its own phase through D-QESA exactly once. Then, the algorithm switches to a random  adaptive beamforming algorithm, BioRARSA2, which we proposed in our previous work \cite{16}. Detailed definitions of parameters of BioRARSA2 can be found in Table 1 of \cite{16}. The pseudocode of the resulting hybrid-QESA is described in Algorithm 4. The Hybrid-QESA combines the advantages of both rapid convergence of deterministic adaptive beamforming algorithms and the strong resistance to dramatic environments of random adaptive beamforming algorithms. This will be demonstrated in our numerical experiments. 

\begin{algorithm}[h] \label{alg: DDSA for static channels for same a}
\caption{Hybrid-Quadratic Equation Search Algorithm} 
Initialize parameters

$\textbf{repeat}$

~~~~$\textbf{for}$ $i\leftarrow 1$ to $N_s$ $\textbf{do}$

~~~~~~~~$\textbf{if}$ $n\leq 3N$ $\textbf{then}$

~~~~~~~~~~~~$\text{test}\leftarrow 0$

~~~~~~~~~~~~$n\leftarrow n+1$

~~~~~~~~~~~~$\theta_{temp}\leftarrow \theta_{cur,i}$

~~~~~~~~~~~~$M_{1}\leftarrow \rss_{temp}(\theta_{temp})$

~~~~~~~~~~~~$\rss[n]\leftarrow M_1$

~~~~~~~~~~~~$n\leftarrow n+1$

~~~~~~~~~~~~$\theta_{temp}\leftarrow \theta_{cur,i}+\alpha$

~~~~~~~~~~~~$M_{2}\leftarrow \rss_{temp}(\theta_{temp})$

~~~~~~~~~~~~$\textbf{if}$ $M_{2}>\rss[n-1]$ $\textbf{then}$ 

~~~~~~~~~~~~~~~~$\rss[n]\leftarrow M_{2}$

~~~~~~~~~~~~~~~~$\theta_{cur,i}\leftarrow \theta_{cur,i}+\alpha$

~~~~~~~~~~~~~~~~$\text{test}\leftarrow 1$

~~~~~~~~~~~~~~~~receiver feeds back a 1 bit to transmitter

~~~~~~~~~~~~$\textbf{else}$

~~~~~~~~~~~~~~~~$\rss[n]\leftarrow \rss[n-1]$

~~~~~~~~~~~~~~~~receiver feeds back a 0 bit to transmitter

~~~~~~~~~~~~$\textbf{end if}$

~~~~~~~~~~~~$n\leftarrow n+1$

~~~~~~~~~~~~$\textbf{if}$ $\text{test}= 1$ $\textbf{then}$ 

~~~~~~~~~~~~~~~~$\theta_{temp}\leftarrow \theta_{cur,i}+\eta-\alpha$

~~~~~~~~~~~~$\textbf{else}$

~~~~~~~~~~~~~~~~$\theta_{temp}\leftarrow \theta_{cur,i}+\eta$

~~~~~~~~~~~~$\textbf{end if}$

~~~~~~~~~~~~$M_3 \leftarrow \rss_{temp}(\theta_{temp})$

~~~~~~~~~~~~$x\leftarrow(M_{1}^2+M_{2}^2)/2$

~~~~~~~~~~~~$\beta_i\leftarrow\arctan\left[ \dfrac{(M_3^2-x)}{(M_2^2-x)}\right]$

~~~~~~~~~~~~$\theta_{cur,i}\leftarrow \theta_{cur,i}+\beta_i$

~~~~~~~~$\textbf{else}$
\end{algorithm}

\begin{algorithm}[h] 
~~~~~~~~~~~~$\textbf{for}$ $j\leftarrow 1$ to L.Helds $\textbf{do}$

~~~~~~~~~~~~~~~~$\mathbf{\delta}$ $\sim uni([-\Delta_k,\Delta_k]^{N_s})$

~~~~~~~~~~~~~~~~$\textbf{if}$ $\rss(\theta_{cur}+\delta)<\rss[n] ~\textbf{then}$

~~~~~~~~~~~~~~~~~~~~$\delta=-\delta$

~~~~~~~~~~~~~~~~$\textbf{end if}$

~~~~~~~~~~~~~~~~$\textbf{if}$ $\rss(\theta_{cur}+\delta)>\rss[n] ~\textbf{then}$

~~~~~~~~~~~~~~~~~~~~$N_T \leftarrow 0$

~~~~~~~~~~~~~~~~~~~~$\omega_n \leftarrow 1$

~~~~~~~~~~~~~~~~~~~~$\textbf{repeat}$

~~~~~~~~~~~~~~~~~~~~~~~~$n \leftarrow n+1$

~~~~~~~~~~~~~~~~~~~~~~~~$\theta_{cur} \leftarrow \theta_{cur}+\delta$

~~~~~~~~~~~~~~~~~~~~~~~~$\rss[n] \leftarrow \rss(\theta_{cur})$

~~~~~~~~~~~~~~~~~~~~~~~~$\omega_n \leftarrow \omega_n+1$

~~~~~~~~~~~~~~~~~~~~$\textbf{untill}$ $\rss(\theta_{cur}+\theta)<\rss[n]$ or $\omega_n>L.Swim$

~~~~~~~~~~~~$\textbf{else}$ 

~~~~~~~~~~~~~~~~$N_T \leftarrow N_T+1$

~~~~~~~~~~~~~~~~~~~~$\textbf{end if}$

~~~~~~~~~~~~~~~~~~~~~~~~$j \leftarrow j+1$

~~~~~~~~~~~~$\textbf{end for}$ 

~~~~~~~~~~~~$\text{Avg. Swim} \leftarrow \max(\rho, \sum_{j-L.Held+1}^j \frac{\omega_n}{L.Helds})$ 

~~~~~~~~~~~~$k \leftarrow k+1$ 

~~~~~~~~~~~~$\Delta_k \leftarrow \Delta_{k-1} \cdot$ Avg. Swim 

~~~~~~~~~~~~$\textbf{if}$ $N_T>L_T$ \textbf{then} 

~~~~~~~~~~~~~~~~$\Delta_k \leftarrow \Delta_{rst}$

~~~~~~~~~~~~~~~~$n \leftarrow n+1$

~~~~~~~~~~~~~~~~$\textbf{if}$ $\rss(\theta_{cur}) \geq \rss[n] ~\textbf{then}$

~~~~~~~~~~~~~~~~~~~~$\rss[n] \leftarrow \rss[n-1]$

~~~~~~~~~~~~~~~~$\textbf{else}$

~~~~~~~~~~~~~~~~~~~~$\rss[n] \leftarrow \rho_T*\rss(\theta_{cur})$

~~~~~~~~~~~~~~~~$\textbf{end if}$

~~~~~~~~~~~~~~~~~~~~$N_T \leftarrow 0$

~~~~~~~~~~~~$\textbf{end if}$

~~~~~~~~$\textbf{end if}$

~~~~$\textbf{end for}$

$\textbf{until}$ stopping criteria reached

\end{algorithm}

\section{Performance analysis}

In this section, we analyze the convergence behavior of our proposed energy-efficient deterministic adaptive beamforming algorithms, including DBSA, D-QESA, D-QESA-E, and hybrid-QESA. We are able to provide rigorous proofs of convergence for all the above algorithms. Specifically, we can guarantee that all our proposed algorithms converge to the global optimal solution.

\subsection{Convergence analysis of DBSA}

The convergence behavior of DBSA is analyzed in the following Theorem. We are able to show that DBSA converges to the global optimal solution irrespective to the initialization parameters.

\begin{Theorem}
For the RSS function defined in (\ref{eq:mag}), let $\left\lbrace \boldsymbol{\theta} [n] \right\rbrace_{n=1}^{\infty}$ be the sequence generated by D-QESA as described by Algorithm 1, where $\boldsymbol{\theta} [n]=\left[ \theta_1[n],~\theta_2[n], \cdots ,~\theta_{N_s}[n] \right]^t$. Then, the resulting sequence converges to the global optimal solution, i.e., $\lim_{n\rightarrow \infty}\rss(\theta [n])=\rss_{max}=\big|\sum_{i=1}^{N_s}a_i\big|.$
\end{Theorem}

\textsl{Proof:}
We design DBSA such that the phase for the $i$-th transmitter is only updated when the RSS function improves. This implies that the RSS function values achieved by the sequence $\left\lbrace \boldsymbol{\theta} [n] \right\rbrace_{n=1}^{\infty}$ are  monotonically non-decreasing. Furthermore, it is clear that $\rss(\boldsymbol{\theta}[n]) \leq \rss_{max} = \big|\sum_ia_i\big|$. Since $\rss(\boldsymbol{\theta}[n])$ is upper bounded and monotonically non-decreasing, the convergence of DBSA is guaranteed by Monotone Convergence Theorem. However, this only guarantees that DBSA converges to local maxima. Fortunately, for the RSS function, all local maxima are global maxima (See \cite{4} for details). Therefore, we can guarantee that DBSA converges to the global optimal solution. 
\hfill{$\blacksquare$}\\~\\
\subsection{Convergence analysis of D-QESA and D-QESA-E}
Here, we want to show that D-QESA  and D-QESA-E indeed converge to the global optimal solution. Note that we do not need the property that all local maxima are global maxima as in the proof of DBSA. This makes our convergence analysis here applicable to a more general set of problem settings.

\begin{Theorem}
For the RSS function defined in (\ref{eq:mag}), let $\left\lbrace \boldsymbol{\theta} [n] \right\rbrace_{n=1}^{\infty}$ be the sequence generated by D-QESA as described by Algorithm 2, where $\boldsymbol{\theta} [n]=\left[ \theta_1[n],~\theta_2[n], \cdots ,~\theta_{N_s}[n] \right]^t$. Then, the resulting sequence converges to the global optimal solution, i.e., $\lim_{n\rightarrow \infty}\rss(\theta [n])=\rss_{max}=\big|\sum_{i=1}^{N_s}a_i\big|.$
\end{Theorem}

\textsl{Proof:}
From (\ref{eq:Substitution}), we can write the RSS function as 
\begin{equation}
\rss(\boldsymbol{\theta} [n]) = \sqrt{P} \left | \sum_{i=1}^{N_s} a_i[n] e^{j\theta_i[n]} \right |  = \sqrt{P} \left | {r}_i + {t}_i  \right | 
\end{equation}

By triangular inequality, we obtain 
\begin{equation}
\rss(\boldsymbol{\theta} [n]) = \sqrt{P} \left | {r}_i + {t}_i  \right | \leq \sqrt{P} \Big( \big | {r}_i \big|+ \big|{t}_i  \big | \Big)
\end{equation}

The aim of D-QESA is to align the phases of $r_i$ and $t_i$. Therefore, we obtain 
\begin{equation}
\rss(\boldsymbol{\theta} [n+1]) = \sqrt{P} \Big( \big | {r}_i \big|+ \big|{t}_i  \big | \Big)
\end{equation}

This implies that  $\rss(\boldsymbol{\theta}[n])$ is a monotonically nondecreasing sequence with an upper bound of $\rss_{max}$. By Monotone Convergence Theorem, we can guarantee convergence of D-QESA. 

Next, we show that D-QESA indeed converges to the global maximum solution by contraction. 
Assume that D-QESA does not converge to the global maximum solution, there is at least one transmitter, say the $i$-th transmitter, not aligned with others. That is, 
\begin{equation}
\rss(\boldsymbol{\theta} [n]) = \sqrt{P} \left | {r}_i + {t}_i  \right | < \sqrt{P} \Big( \big | {r}_i \big|+ \big|{t}_i  \big | \Big)
\end{equation}

However, it is clear that $\rss(\boldsymbol{\theta} [n])$ can be improved by running D-QESA for the $i$-th transmitter to align the phases of $r_i$ and $t_i$. This contradicts with our original assumption. 
Therefore, D-QESA will not stop at such a point, and it will stop only when 
\begin{equation}
\rss(\boldsymbol{\theta} [n]) = \sqrt{P} \left | {r}_i + {t}_i  \right | = \sqrt{P} \Big( \big | {r}_i \big|+ \big|{t}_i  \big | \Big)
\end{equation}

In other words, D-QESA is guaranteed to converge to the global optimal solution. \hfill{$\blacksquare$}

For the case of equal channel gains, we have the following theorem that describes and analyzes the convergence behavior of D-QESA-E. 
\begin{Theorem}
For the RSS function defined in (\ref{eq:mag}), let $\left\lbrace \boldsymbol{\theta} [n] \right\rbrace_{n=1}^{\infty}$ be the sequence generated by D-QESA-E as described by Algorithm 3, where $\boldsymbol{\theta} [n]=\left[ \theta_1[n],~\theta_2[n], \cdots ,~\theta_{N_s}[n] \right]^t$. Then, the resulting sequence converges to the global optimal solution, i.e., $\lim_{n\rightarrow \infty}\rss(\theta [n])=\rss_{max}=N_s a$.
\end{Theorem}

\textsl{Proof:}
The proof is almost the same at that of Theorem 2 and is not repeated  here.
\hfill{$\blacksquare$}

\subsection{Convergence analysis of Hybrid-QESA}

The convergence analysis of Hybrid-QESA can be done through combining the convergence analysis of D-QESA and BioRASA2. The following theorem makes this statement more precise.

\begin{Theorem}
For the RSS function defined in (\ref{eq:mag}), let $\left\lbrace \boldsymbol{\theta} [n] \right\rbrace_{n=1}^{\infty}$ be the sequence generated by D-QESA as described by Algorithm 2, where $\boldsymbol{\theta} [n]=\left[ \theta_1[n],~\theta_2[n], \cdots ,~\theta_{N_s}[n] \right]^t$. Then, the resulting sequence converges to the global optimal solution, i.e., $\lim_{n\rightarrow \infty}\rss(\theta [n])=\rss_{max}=\big|\sum_{i=1}^{N_s}a_i\big|.$
\end{Theorem}

\textsl{Proof:}
No matter how many rounds we run D-QESA before we switch into BioRASA2, the final phase configuration of D-QESA simply serves as a good initial point for BioRASA2. Since BioRASA2 is guaranteed to converge to the global optimal solution (See \cite{16}), Hybrid-QESA is also guaranteed to converges to the global optimal solution.
\hfill{$\blacksquare$}
\begin{remark}
The proof of convergence for DBSA and Hybrid-QESA depends on the property that all local maxima are global maxima for the RSS function. However, we can guarantee convergence for D-QESA  and D-QESA-E without this property.  This indicates that convergence of D-QESA  and D-QESA-E can be guaranteed for a wider set of problem settings, e.g., when the objective function is not the RSS function.
\end{remark}

\section{Simulation results}
In this section, we provide simulation results that evaluate the efficiency of our proposed deterministic adaptive beamforming algorithms. We set the transmitted symbol power to be $\sqrt P= 1$. In our simulations, all channels realizations are assumed to be zero-mean, unit variance  i.i.d Rayleigh flat fading.  All simulations are obtained with the same number of distributed transmitters, i.e., $N_s = 100$.
Four types of channel and network settings are considered, i.e., noiseless channels, noisy channels, random node addition and removal, and noiseless time-varying channels.

\subsection{Noiseless channels}

\begin{figure}
\begin{center}
\resizebox{.5\textwidth}{!}{\includegraphics{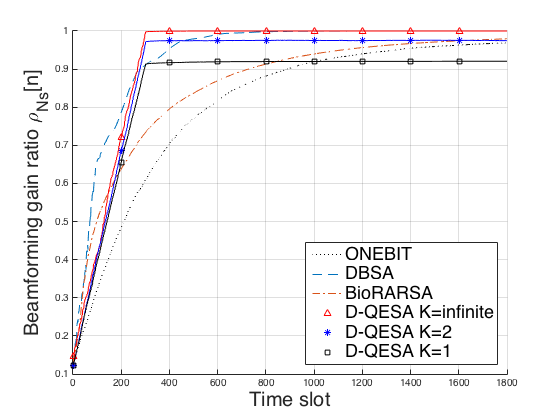}}\label{fig 1}
\caption{ Comparison of the convergence behavior between D-QESA for different number of feedback bits  ($\text{K}=1, 2 , \text{inifite}$) with one-bit random scheme, BioRASA and DBSA for noiseless channels.
}
\end{center}
\end{figure}

\begin{figure}
\begin{center}
\resizebox{.5\textwidth}{!}{\includegraphics{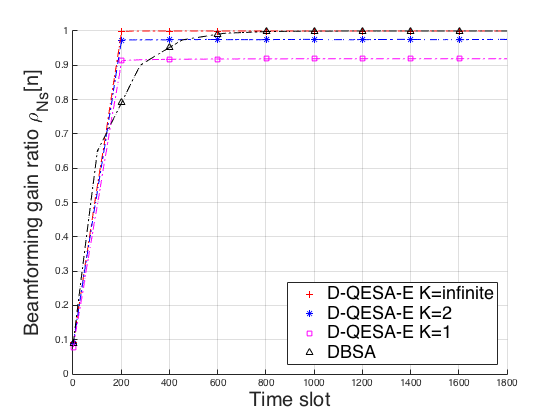}}
\caption{Evolution of D-QESA-E with different bandwidth of the reverse feedback link for the special case of equal channel gains, i.e., $a_i=a,~\forall i$. Evolution of DBSA is also included for comparison.}
\end{center}
\end{figure}

In Figure 1 and Figure 2, we demonstrate how the the beamforming gain ratio evolves for noiseless channels. 
Each simulation curve is obtained by averaging over $100$ randomly generated channel realizations.
In Figure 1, we compare the convergence behavior between DBSA and D-QESA for different number of feedback bits with the following schemes: a) one-bit random scheme proposed in \cite{2}, and b) BioRASA proposed in \cite{6}. We clearly observe that our proposed DBSA and D-QESA exhibit superior convergence behavior over random adaptive beamforming schemes. Comparison between DBSA and D-QESA reveals that better performance is achieved for DBSA if there is only one-bit feedback. When more than $2$ bits are available for feedback, D-QESA becomes a more attractive alternative. We can observe that beamforming gain ratios for D-QESA almost all converge to a fixed value in about $300$ iterations, which accounts for about one round-robin in D-QESA.
That is, one round-robin is enough for the convergence of D-QESA. It is clear that faster convergence can be achieved with more number of feedback bits. This gain, however, diminishes quite fast as the number of feedback bits increases. For example, with 2 bits of feedback, $97\%$ of $\rss_{max}$ can be reached in Figure 1. This indicates that D-QESA can also operate well with limited bandwidth of the reverse feedback link.

In Figure 2, we compare the convergence behavior of D-QESA-E for different bandwidth of the reverse feedback link with DBSA for $a_i=a$. It is clear that beamforming gain ratios almost all converge to a fixed value in about $200$ iterations, which accounts for about one round-robin in D-QESA-E. This demonstrates the rapid convergence of D-QESA-E. Indeed, there is about a $1/3$ reduction in convergence time when compared with D-QESA, for which about $300$ iterations are required before convergence. Furthermore, D-QESA-E begins to outperform DBSA even with one-bit feedback before reaching its converged value of about $91\%$ of $\rss_{max}$.

\subsection{Noisy channels}
\begin{figure}
\begin{center}
\resizebox{.5\textwidth}{!}{\includegraphics{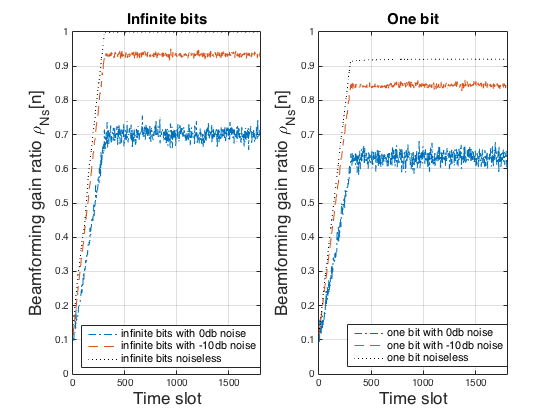}}
\caption{Comparison of the convergence behavior of D-QESA under different noise powers of  $-10$ db and $0$ db, respectively. The influence of D-QESA can be observed for noisy channels with one-bit and infinite number of bits of feedback . 
}
\end{center}
\end{figure}

In Figure 3, we demonstrate the evolutions of D-QESA for noisy channels. Here, the performance curve is not obtained by averaging over $100$ channel realizations since we want to observe the influence of noise. We compare the convergence behavior of D-QESA under different noise powers of  $-10$ db and $0$ db, respectively.  
When the noise power is same as the signal power, i.e., when noise power is $0$ db, the beamforming gain ratio is poor and reaches only about $62\%$ of $\rss_{max}$ for one-bit feedback and $70\%$ for infinite feedback bits.
For a more practical noise level of  $-10$ db, the beamforming gain ratio is a bit lower than that of a noiseless case and reaches  about $85\%$ of $\rss_{max}$ for one-bit feedback and $94\%$ for infinite feedback bits. This demonstrates that D-QESA is still quite robust against noise for most practical scenarios.

\subsection{Random node addition/removal}

\begin{figure}
\begin{center}
\resizebox{.5\textwidth}{!}{\includegraphics{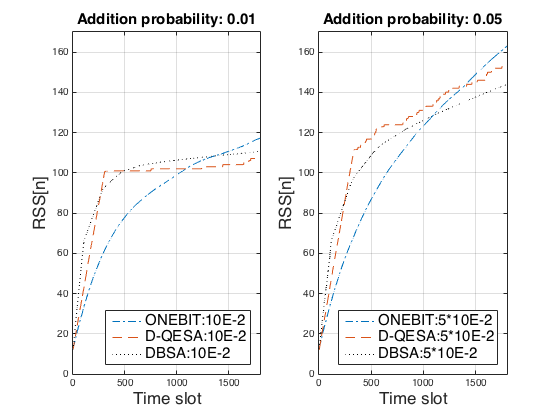}}
\caption{Evolutions of D-QESA and DBSA are compare against the random one-bit scheme. The system with node addition probabilities are set to be 0.01 and 0.05, respectively. 
}
\end{center}
\end{figure}

\begin{figure}
\begin{center}
\resizebox{.5\textwidth}{!}{\includegraphics{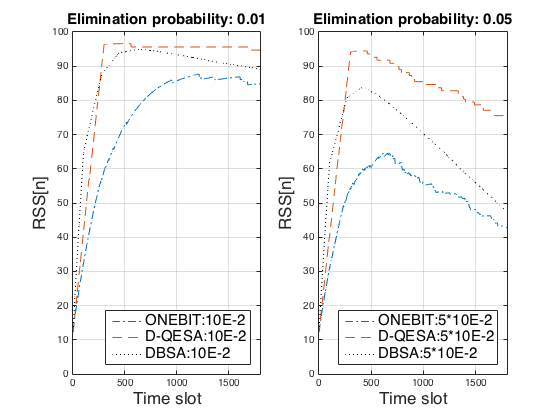}}
\caption{Evolutions of D-QESA and DBSA are compare against the random one-bit scheme. The system with node elimination probabilities are set to be 0.01 and 0.05, respectively. 
}
\end{center}
\end{figure}

Here, we investigate the influence of random node addition and removal on adaptive beamforming algorithms. That is, we want to investigate whether our proposed algorithms are robust against uncertainties in wireless network topology. Note that we observe the RSS function value instead of the beamforming gain ratio here since random node addition and removal alter $\rss_{max}$ such that the beamforming gain ratio may not be a good measure for performance. We choose two node addition and removal probabilities of 0.01 and 0.05, respectively. In Figure 4, we demonstrate the evolution of the RSS function under random node addition. We can see that fast initial convergence is maintained for both D-QESA and DBSA. As we move along further, the one-bit random scheme begins to outperform both D-QESA and DBSA, indicating that random adaptive beamforming algorithms are more robust against random node addition for wireless sensor/relay networks. For the case where the random node addition probability is set to be $0.05$, D-QESA outperforms DBSA slightly. This suggest that D-QESA might be more robust against network uncertainty.

In Figure 5, we demonstrate the evolution of the RSS function under different random node removal probabilities. A quite different phenomenon can be observed. It is clear that D-QESA and DBSA outperforms the one-bit scheme quite significantly. 
Furthermore, D-QESA demonstrates much superior performance than the other two algorithms, especially when the node removal probability is higher. Specifically, D-QESA is able to maintain its RSS function value to be above $90\%$ and $70\%$ of $\rss_{max}$ for removal probabilities of $0.01$ and $0.05$, respectively.
This is a surprising result since random adaptive beamforming algorithms are believed to be more robust against such uncertainties in general. 
This numerical experiment demonstrates that D-QESA is not only an attractive adaptive beamforming algorithm for static channels but also a robust algorithm against uncertainties in wireless network topology. 
This is a more desirable feature since random node removal is more detrimental for distributed beamforming schemes since there is a sudden decrease in the RSS function value.

\subsection{Time-varying channels}

\begin{figure}
\begin{center}
\resizebox{.5\textwidth}{!}{\includegraphics{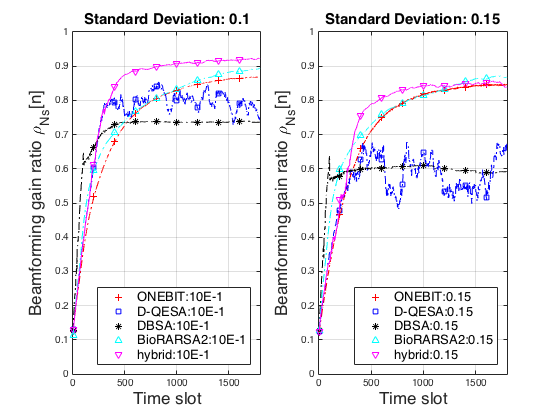}}
\caption{Comparison of the beamforming gain ratio between Hybrid-QESA, D-QESA, DBSA, the one-bit random scheme, BioRASA, and BioRASA2 for different standard deviations (0.1 and 0.15) of the channel variation under time-varying channels. 
}
\end{center}
\end{figure}

In Figure 6, we consider the case of time-varying channels described by (\ref{time-varying}), where the standard deviation of 
$\xi_i[n]$ is set to be $0.1$ and $0.15$, respectively. Here, we compare the convergence behavior of D-QESA, DBSA, and hybrid-QESA with the one-bit random scheme, BioRASA, and BioRASA2. 
When the standard deviation $\delta_{\xi_i}$ is set to be $0.1$, we can see that the random adaptive beamforming algorithms have better performance than their deterministic counterparts. D-QESA still achieves a decent RSS function value but experiences severe fluctuation under time-varying channels. Hybrid-QESA combines the benefits of both types of adaptive beamforming algorithms and achieves the best performance over all the rest. Compared with BioRASA2, it can be observed that a fast initial convergence separates the performance of hybrid-QESA from BioRASA2. 
When $\delta_{\xi_i}=0.15$, hybrid-QESA still outperforms most schemes. BioRASA2 has a slight edge in performance as we move further along. This suggests that random adaptive beamforming algorithms are still attractive alternatives under severe channel variations. However, strong performance for hybrid-QESA can be observed for both channel settings.

\section{Conclusion}
In this study, we proposed energy-efficient deterministic adaptive beamforming algorithms for distributed sensor/relay networks. Specifically, we proposed DBSA, D-QESA, D-QESA-E, and a hybrid algorithm, hybrid-QESA, that combines the benefits of both deterministic and random adaptive beamforming algorithms. DBSA is inspired by the idea of bisection search. D-QESA is inspired by a shift of perspective that views each transmission as a RSS function evaluation and views the problem of adaptive beamforming for each distributed transmitter as solving quadratic equations with independent variables. 
We further provided rigorous convergence analysis for all our proposed algorithms and proved that the global optimal solution is reached for all our proposed algorithms. In our numerical experiments, we demonstrated that superior performance is achieved by our proposed DBSA and D-QESA over random adaptive beamforming algorithms for static channels. Surprisingly, D-QESA is also more robust against random node removal than random adaptive beamforming algorithms. For time-varying channels, hybrid-QESA indeed achieves the best performance since it combines the benefits of both types of adaptive beamforming algorithms. In summary, our proposed deterministic algorithms demonstrate superior performance both in terms of convergence time and robustness against channel and network uncertainties. This surprising new finding challenges the conventional belief that deterministic algorithms are faster in convergence while random algorithms are more robust against uncertainties. We hope that this work can generate more interests  in the studies of both deterministic and random adaptive beamforming algorithms.

\bibliographystyle{IEEEtran}
\bibliography{./reference}
\end{document}